\documentstyle[prd,aps,psfig,twocolumn,floats]{revtex}
\newcommand{\cd}{\cdot}

\newcommand{\de}{\delta}
\newcommand{\De}{\Delta}
\newcommand{\ep}{\epsilon}

\newcommand{\la}{\lambda}

\newcommand{\si}{\sigma}

\newcommand{\Th}{\Theta}

\newcommand{\ra}{\rightarrow}
\newcommand{\bm}[1]{\mbox{\boldmath $#1$}}

\newcommand{\be}{\begin{equation}}
\newcommand{\ee}{\end{equation}}
\newcommand{\bea}{\begin{eqnarray}}
\newcommand{\eea}{\end{eqnarray}}
\newcommand{\bean}{\begin{eqnarray*}}
\newcommand{\eean}{\end{eqnarray*}}

\newcommand{\lsim}{\stackrel{<}{\sim}}
\newcommand{\lan}{\langle}
\newcommand{\ran}{\rangle}

\title{Cosmic Microwave Background Anisotropies
 from Scaling Seeds: Generic Properties of the Correlation Functions}
\author{R. Durrer and M. Kunz}
\address{D\'epartement de Physique Th\'eorique,
	 Universit\'e de Gen\`eve,
	24 quai Ernest Ansermet, CH-1211 Gen\`eve 4, Switzerland}
\begin{document}
\maketitle
\begin{abstract}
In this work we present a partially new method to analyze fluctuations
which are induced by causal scaling seeds. We show that the power
spectra due to this kind of seed perturbations are determined by
five analytic functions, which we determine numerically for a special
example. We put forward the view that, even if recent work disfavors
the models with cosmic strings and global $O(4)$ texture, causal
scaling seed perturbations merit a more thorough and general analysis,
which we initiate in this paper.
\end{abstract}

\section{Introduction}
At present, two ideas to explain the origin of large scale structure in 
the universe are primarily investigated: 
\begin{itemize}
\item
Perturbations in the energy 
density of matter may have emerged from quantum fluctuations, which,   
during a phase of inflationary expansion, are  stretched beyond the Hubble 
scale and 'freeze in' as classical fluctuations in the energy density of 
the cosmic fluid\cite{BST}. These small initial fluctuations then 
evolve according to linear cosmological perturbation theory.
\item
A phase transition in the early universe may lead to topological defects 
which seed fluctuations in the cosmic fluid \cite{Ki}. In this case, 
the fluid fluctuations
vanish initially and evolve according to inhomogeneous
cosmic perturbation equations. The stress energy of the topological defects 
plays the role of the source term.
In order for the gravitational field of the defects to be sufficiently 
strong to seed cosmic structure, we have to require $\ep=4\pi G\eta^2\sim 
10^{-5}$, where $\eta$ denotes the energy scale of the phase transition. This
yields $\eta\sim 10^{16}$GeV, a typical GUT scale. Topological defects 
which neither over-close the universe nor die out are either cosmic 
strings or global defects.
\end{itemize}
The anisotropies in the cosmic microwave background (CMB) provide an 
important tool to discriminate between different models. On very large angular
scales both classes of models lead to a Harrison Zel'dovich spectrum of
fluctuations. For inflationary models this can easily be derived 
analytically. For defect models, the spectra were originally found numerically
\cite{PST,DZ,Shellard}. Analytical arguments for this behavior are
given in \cite{roma}. On intermediate scales, inflationary models predict
 a series of peaks  due to acoustic oscillations in the baryon/photon 
fluid prior to recombination\cite{BE}. Present observations seem to 
confirm these peaks even though the error bars are still 
considerable\cite{obs}.
 
Recently, several investigations led to the conclusion that cosmic strings
\cite{Shel,Andy} and global ${\cal O}(N)$ defects \cite{DGS,Turok,DKLS} 
do not reproduce the acoustic peaks indicated by present data. This led
\cite{Turok} and \cite{Andy} to the conclusion that models of cosmic 
structure formation with scaling causal defects are ruled out. 

However, in a very  simple parameterization of two families of 
more general scaling causal seed models,
we were able to fit present data very well\cite{DKLS}.
We are thus convinced that it is too early yet to completely abandon scaling 
causal seeds as a mechanism for structure formation. In contrary, we think
it is extremely useful to study  them  in full generality,
ignoring in a first step the physical origin of the seeds. This purely
phenomenological point of view is analogous to inflationary models 
where one sometimes manufactures the inflationary potential
to yield the required spectrum of initial perturbations.  

To determine  the power spectrum of the radiation and 
matter perturbations induced by seeds, we need to know the
two point correlation functions of the seed energy momentum tensor.
In this paper we present a simple parameterization of these functions
and point out an error frequently made in the literature. We then exemplify
our findings with the large $N$ limit of global scalar fields \cite{TS,KD}.

For simplicity, we work in a spatially flat Friedman universe. The metric is thus
given by 
\[ds^2=a(t)^2(dt^2-\de_{ij}dx^idx^j)~,\]
 where $t$ denotes conformal time.

\section{Correlation functions of causal scaling  seeds}
We define seeds to be any   non-uniformly distributed form of energy, which 
contributes only a small fraction to the total energy density of the universe
and which interacts with the cosmic fluid only gravitationally.

We parameterize the energy momentum tensor of the seed by
\be
 T^{(seed)}_{\mu\nu} = M^2\Th_{\mu\nu} ~,
\ee
where $M$ denotes a typical energy of the seed and $\Th_{\mu\nu}$ is a 
random variable. We assume that ensemble averages and and spatial 
averaging coincide, the usual ergodicity hypothesis. Furthermore, we 
assume the random process $\Th_{\mu\nu}$ to be spatially homogeneous and 
isotropic, so that the two point correlation function
\bea
\lefteqn{\lan \Th_{\mu\nu}({\bf x},t)\Th_{\la\rho}({\bf x+y},t')\ran
	=} \nonumber\\  &&
 {1\over V}\int
    \Th_{\mu\nu}({\bf x},t)\Th_{\la\rho}({\bf x+y},t')d^3x \equiv 
	C_{\mu\nu\la\rho}({\bf y},t,t')
\eea
is a function of the difference $\bf y$ only. Due to the expansion of 
the universe, which breaks time translation symmetry, we however expect
$C$ to depend on $t$ and $t'$ and not just on the difference $t-t'$.
We consider causal seeds. Causality requires

\be C_{\nu\mu\la\rho}({\bf y},t) = 0, \mbox{ if }~~|{\bf y}| > t+t'~.\ee
The two point function in position space thus has compact support 
which implies that  its Fourier transform is analytic.

We define a seed to be scaling, if the Fourier transform,
\be
\widehat{C}_{\mu\nu\la\rho}({\bf k},t,t') = \lan 
	\widehat{\Th}_{\mu\nu}({\bf k},t)
	\widehat{\Th}^*_{\la\rho}({\bf k},t')\ran ~,
\ee
contains no  dimensional parameter other than $t,t'$ and
$k$\footnote{We neglect the transition from a radiation to a
matter dominated universe, which actually breaks scaling, since the
scale $t_{eq}$, the time of equal matter and radiation density enters
the problem. In numerical examples, we have found that this transition
in general leads to somewhat different  decay laws for the correlation
functions at large $kt$, but it will not alter our main conclusions}.
This implies that
the Ricci curvature induced by the seeds is a function of $k$ and $t$ only,
multiplied by the dimensionless parameter $\ep=4\pi GM^2$.
Since we define Fourier transforms with the normalization
\bea 
\hat{f}({\bf k})&=&{1\over \sqrt{V}}\int d^3x f({\bf x})\exp(i{\bf kx}) ~;\\
	f({\bf x}) &=& {\sqrt{V}\over (2\pi)^3}\int d^3k \hat{f}({\bf k})
	\exp(-i{\bf kx})~,
\eea
and since $\Th_{\mu\nu}({\bf x},t)$ has the dimension of (length)$^{-2}$,
$\widehat{C}$ has the dimension of an inverse length. From scaling we 
therefore conclude that for purely dimensional reasons, we can write the
correlations functions in the form
\be
\hat{C}_{\mu\nu\la\rho}({\bf k},t,t') =
	{1\over\sqrt{tt'}}F_{\mu\nu\la\rho}(\sqrt{tt'}\cd{\bf k},t'/t)~,
\ee
where $F_{\mu\nu\la\rho}$ is a dimensionless  function of the four variables
 $z_i\equiv\sqrt{t't} k_i$ and $r\equiv t'/t$, which is analytic in $z_i$.

We also require the seed to decay inside the horizon, which
implies
\be \hat{C}_{\mu\nu\la\rho}({\bf k},t,t')\ra_{kt\ra \infty}= 
	\hat{C}_{\mu\nu\la\rho}({\bf k},t,t')\ra_{kt'\ra \infty}= 0~.
\label{decay}\ee
Furthermore,
since the seeds interact with the cosmic fluid only gravitationally, $\Th$ 
satisfies covariant energy momentum conservation,
\be \Th^{\mu\nu};_\nu =0 ~. \ee
With the help of these four equations, we can, for example, express the 
temporal components, $\Th_{0\mu}$ in terms of the spatial ones, $\Th_{ij}$.
The seed correlations are thus fully determined by the spatial 
correlation  functions $\widehat{C}_{ijlm}$. Statistical isotropy, scaling and
symmetry in $i,j$ and $l,m$ as well as under the transformation 
$i,j;k;t \ra l,m;-k;t'$
require the following form for the spatial correlation functions:
\bea
\lefteqn{\widehat{C}_{ijlm}({\bf k},t,t') =} \nonumber \\  
	&&{1\over \sqrt{tt'}}
	[z_iz_jz_lz_mF_1(z^2,r) + \nonumber \\  &&
  (z_iz_l\de_{jm}+z_iz_m\de_{jl}+z_jz_l\de_{im}
 +z_jz_m\de_{il})F_2(z^2,r) + \nonumber \\ &&
z_iz_j\de_{lm}F_3(z^2,r)/r +z_lz_m\de_{ij}F_3(z^2,1/r)r +
\nonumber\\ &&
+\de_{ij}\de_{lm}F_4(z^2,r) +
(\de_{il}\de_{jm}+\de_{im}\de_{jl})F_5(z^2,r)] ~,
\label{Cijlmansatz}
\eea
where the functions $F_a$ are  analytical functions of 
$z^2\equiv tt'k^2$, and for  $ a \neq 3$ they are
invariant under the transformation $r\ra 1/r$, $F_a(z^2,r)=F_a(z^2,1/r)$.
The positivity of the power spectra 
$\hat{C}_{ijij}({\bf k},t,t)=\lan|\Th_{ij}|^2\ran$ leads to a series of positivity 
conditions for the functions $F_a$:
\bea
0 &\le& F_5(z^2,1) ~,  \label{pF5}\\
0 &\le& F_4(z^2,1)+2F_5(z^2,1) ~, \label{p2}\\
0 &\le& z^2F_2(z^2,1)+F_5(z^2,1) ~,\\
0 &\le& z^4F_1(z^2,1)+4z^2F_2(z^2,1)+3F_5(z^2,1)~,\\
0 &\le& z^4F_1(z^2,1)+2z^2(F_3(z^2,1)+2F_2(z^2,1)) \nonumber\\
	&& +F_4(z^2,1)+2F_5(z^2,1)~. \label{p5}\eea
\vspace{0.5cm}
Since $\hat{C}_{ijlm}$ is the Fourier transform of a real
function, 
\be
\hat{C}_{ijlm}({\bf k},t,t')^* =\hat{C}_{ijlm}(-{\bf k},t,t')~, 
\ee
and thus the 
ansatz (\ref{Cijlmansatz}) implies that the functions $F_a(z^2,r)$ are
real. Furthermore, decay inside the horizon (condition (\ref{decay})) yields
\be
 \lim_{z^2r\ra\infty}F_a(z^2,r)=\lim_{z^2/r\ra\infty}F_a(z^2,r)=0~.
\label{lim} \ee
In addition, analyticity implies that the functions $F_a$ do not
diverge in the limit $z\ra 0$, thus 
\[\lim_{z\ra 0}F_a(z,r)=A_a(r)   \]
with 
\[ A_a(r)=A_a(1/r)
	~\mbox{ for all } a\neq 3~. \]

As an example, we have worked out the functions $F_1$ to $F_5$ in 
the large $N$ limit of global scalar field seeds. A discussion of this
simple model of scaling causal seeds ands its relation to the texture
model of structure formation can be found in \cite{TS} and  \cite{KD} .
In Figs.~1 and 2 we present the functions  $F_5(z^2,r)$ and $F_2(z^2,r)$.
\begin{figure}[h]
\centerline {\psfig{figure=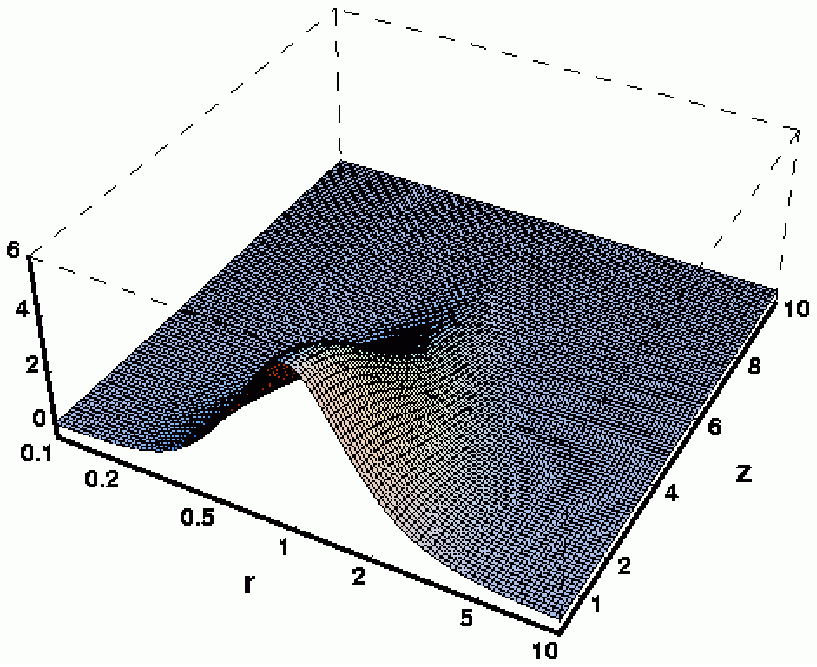,width=7.5cm}}
\centerline {\psfig{figure=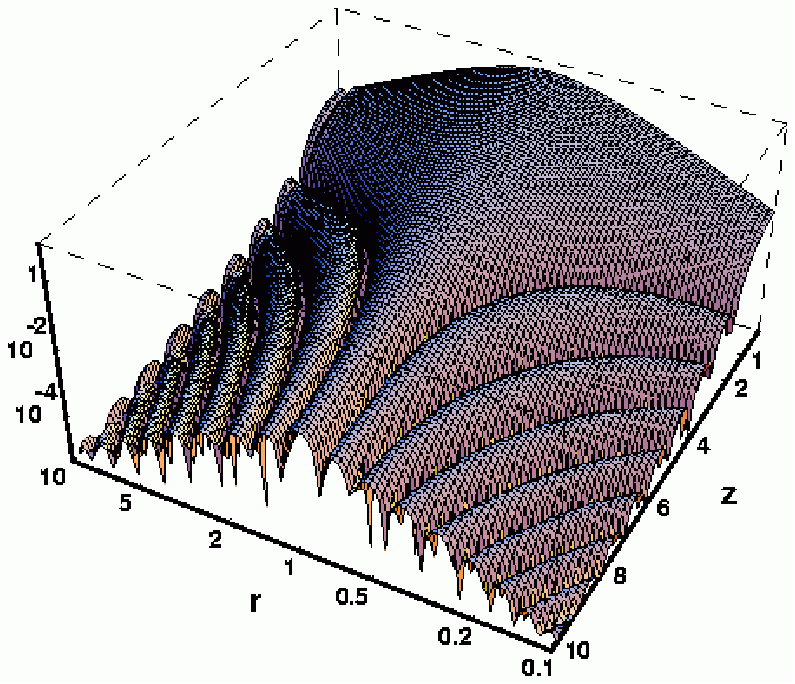,width=7.5cm}}
\caption{The function $F_5(z,r)$, linear (top) and $|F_5(z,r)|$
logarithmic are shown. Because of their small amplitude, the oscillations are
virtually invisible in the linear plot.
To show the symmetry $r\ra 1/r$, the $r$-axis is chosen logarithmic in
both plots.}
\label {fig1}
\end{figure}
\begin{figure}[ht]
\centerline {\psfig{figure=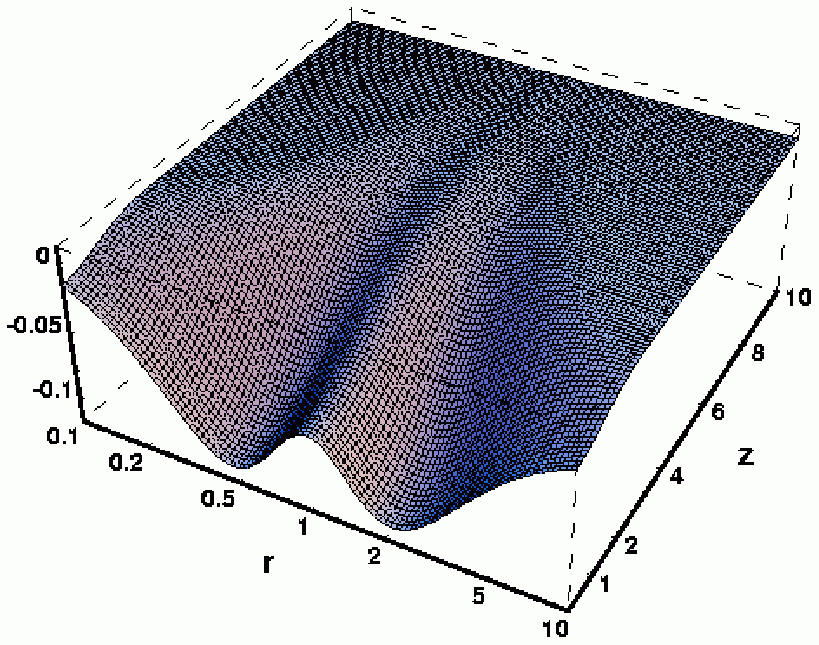,width=7.5cm}}
\centerline {\psfig{figure=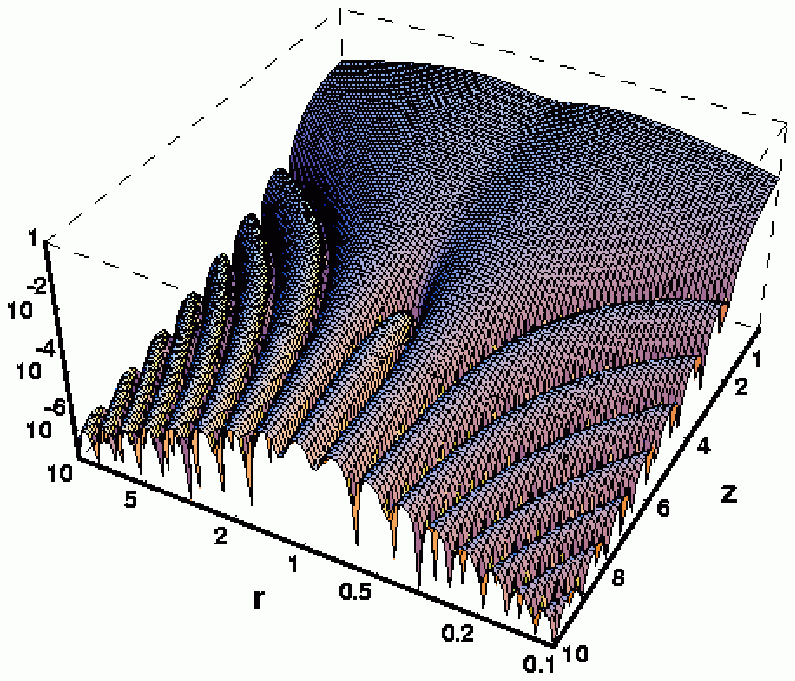,width=7.5cm}}
\caption{The same as Fig.~1 for the function $F_2(z,r)$.}
\label {fig2}
\end{figure}

The symmetry under the transition $r\ra 1/r$ is clearly visible. Also
the conditions that $F_a\ra 0$ if either $z\ra \infty$ or $r\ra 0$ or 
$r\ra \infty$ which follows from Eq.~(\ref{lim}) is evidently satisfied. 
For fixed $z$ the functions oscillate with a frequency which grows
with $z$. Since the amplitude decays rapidly, these oscillations are
only visible in the log-plots. The correlations always decay like
power laws, never exponentially.
 
The equal time correlation functions, $F_1(z^2,1)$ to $F_5(z^2,1)$ 
are plotted in Fig.~3. In Fig.~4 we show $A_a(r)$.

\begin{figure}[ht]
\centerline {\psfig{figure=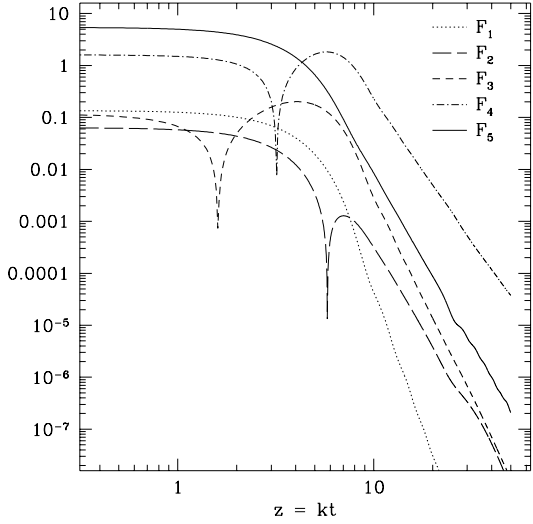,width=7.5cm}}
\caption{The functions $|F_i(z,1)|$ are shown. The zeros are visible
as spikes in the log-plot. (Further below, at $z\sim 30$, also $F_1$
passes through zero.)}
\label {fig3}
\end{figure}
\begin{figure}[ht]
\centerline {\psfig{figure=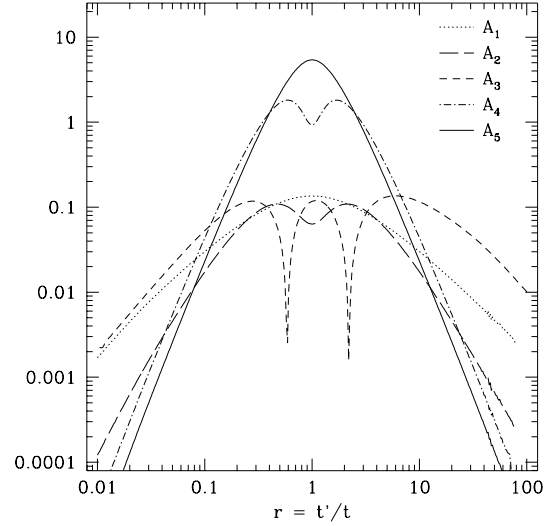,width=7.5cm}}
\caption{The functions $|A_i(r)|=|F_i(0,r)|$ are shown. As discussed
in the text, all of them except $A_3$ are symmetrical under the
transition $r\ra 1/r$.}
\label {fig5}
\end{figure}
All the functions, except $F_5$ which is constrained by Eq.~(\ref{pF5})
go through $0$ (For $F_1$ the passage through $0$ is not visible 
since it occurs only
at $z\simeq 30$). We have also verified the positivity constraints
Eq.~(\ref{p2}) to Eq.~(\ref{p5}). The asymptotic behavior of the 
functions can be obtained analytically.
The same is true for the functions $A_1$ to $A_5$. As argued
above, all the functions $A_i$ except $A_3$ are symmetric under $r\ra
1/r$. The techniques employed to calculate the functions $F_i$ are
analogous to those explained in \cite{KD}. 

\section{Scalar, vector and tensor decomposition and  CMB anisotropies}
The energy momentum tensor of seeds is often split into scalar, vector 
and tensor perturbations, since the time evolution of each of these 
components is independent. Furthermore, due to statistical isotropy,
the scalar, vector and tensor modes are uncorrelated.

A suitable parameterization of this decomposition is
\be \Th_{ij} = \de_{ij}f_p -(k_ik_j-{k^2\over 3}\de_{ij})f_\pi +
	{1\over 2}(w_ik_j+w_jk_i) +\tau_{ij}  ~, \label{deco} \ee
where $f_p,~f_\pi$ are arbitrary random functions of {\bf k}; {\bf w} 
is a transverse vector, ${\bf w\cd k}=0$; and $\tau_{ij}$ is a 
symmetric, traceless, transverse tensor, $\tau^i_i=\tau_{ij}k^j=0$.
The variables $f{\bf .}~, {\bf w}$ and $\tau_{ij}$ represent the 
scalar, vector and tensor degrees of freedom of $\Th_{ij}$.
The functions $F_1$ to $F_5$ determine the correlations:
To work out the correlation functions we use
\bea
 f_p &=& {1\over 3}\Th_i^i ={1\over 3}\de_{ij}\Th^{ij}~, \label{fp}\\
f_\pi &=& -{3\over 2k^2}(\hat{k}^i\hat{k}^j-{1\over 3}\de_{ij})\Th_{ij}
  = {3\over 2k^2}S_{ij}\Th^{ij}~, \label{fpi}\\
w_i &=& {2\over k}(\de_i^{~l}\hat{k}^m-\hat{k}_i\hat{k}^l\hat{k}^m)\Th_{lm}
	= V_i^{~lm}\Th_{lm}~, \label{w}\\
\tau_{ij} &=&(P_{il}P_{jm}-(1/2)P_{ij}P_{lm})P^{ma}P^{lb}\Th_{ab}
\nonumber\\
	&=&  T_{ij}^{~~ab}\Th_{ab}~,\label{tau}\\
\mbox{ where }&& P_{ij}=\de_{ij}-\hat{k}_i\hat{k}_j ~~~,~~~ 
	\hat{k}^i =k^i/k  \label{proj}
\eea
$P_i^{~j}$ is the projection operator onto the space orthogonal 
to {\bf k} and $S_{ij}$,
$V_i^{~lm}$ and $T_{ij}^{~~ab}$ are the projection operators to the scalar
vector and tensor parts of $\Th_{ij}$.

Using these identities and our ansatz (\ref{Cijlmansatz}), one easily 
verifies
\bea
\lan f_p(t)w_i^*(t')\ran &=&\lan f_\pi(t) w_i^*(t')\ran = 0\\
	\lan f_p(t)\tau_{ij}^*(t')\ran &=& 
 \lan f_pi(t) \tau_{ij}^*(t')\ran = 0 \label{C0s} \\
\lan w_i(t)\tau_{jl}(t')\ran 	&=& 0 \label{C0vt}\\
\lan f_p(t)f_p^*(t')\ran &=& {1\over 3\sqrt{tt'}}[2F_5(z^2,r)+3F_4(z^2,r)
  \nonumber \\ && + z^2(F_3(z^2,r)/r+F_3(z^2,1/r)r) + \nonumber \\
  && {4\over 3}z^2F_2(z^2,r) + {1\over 3}z^4F_1(z^2,r)]  \label{Cp}\\
\lan f_\pi(t)f_\pi^*(t')\ran &=& {1\over\sqrt{tt'}k^4}[3F_5+4z^2F_2
	+ z^4F_1]  \label{Cpi} 
\eea
\bea
\lefteqn{\lan f_p(t)f_\pi^*(t')\ran =} \nonumber\\
 &&   -\sqrt{tt'}[ {1\over 3}z^2F_1(z^2,r)
	+{4\over 3}F_2(z^2,r) 
	+rF_3(z^2,1/r)] \label{Cppi}\\
\lefteqn{\lan w_i(t)w_j^*(t')\ran =} \nonumber\\
&&  {4\over k^4\sqrt{tt'}}[F_5+z^2F_2]
	(k^2\de_{ij}-k_ik_j) \label{Cw}\\
\lefteqn{\lan\tau_{ij}(t)\tau^*_{lm}(t')\ran =} \nonumber \\
&&	{1\over\sqrt{tt'}}F_5[\de_{il}\de_{jm}
+\de_{im}\de_{jl}   -\de_{ij}\de_{lm} + 
	k^{-2}(\de_{ij}k_lk_m + \nonumber \\ &&
	\de_{lm}k_ik_j -\de_{il}k_jk_m - \de_{im}k_lk_j -\de_{jl}k_ik_m
	-\de_{jm}k_lk_i) + \nonumber \\
&&	k^{-4}k_ik_jk_lk_m] \label{Ctau}
    ~.\eea
It is interesting to note that although $\widehat{C}_{ijlm}$ is analytic, the 
correlation functions of the scalar vector and tensor components, in 
general, are not. The reason for that is that the projection operators
$S, V$ and $T$ are not analytic. This is important. It implies, 
{\em e.g.}, that the anisotropic stresses in general have a white noise
 and not a $k^4$ spectrum as  erroneously concluded in \cite{Joao,Sper,Nat}. 
The scalar 
anisotropic stress potential thus diverges on large scales,
$\lan |f_\pi|^2\ran \propto 1/(tk^4)$ for $kt\ll 1$. A result which we
also have obtained in the large-$N$ limit and in numerical simulations
of $O(N)$ models. The power spectrum of the
scalar anisotropic stress potential $f_\pi$ is analytic if and only if
vector and tensor perturbations are absent, $F_5=F_2=0$.
In the generic situation, $F_5(z=0,r=1)=A_5(1)\neq 0$. We thus expect 
the following
relation between scalar vector and tensor perturbations of the 
gravitational field on super-horizon 
scales, $x\equiv kt\ll 1$: (The equations for the scalar, vector and tensor
gravitational potentials  in terms of $f.$, {\bf w} and
$\tau_{ij}$ can be found in \cite{d90} and \cite{RuthReview}.) 
\bea
\lan |\Phi-\Psi|^2\ran &\sim& {12\ep^2\over tk^4}A_5(1)  \\
\lan |\si_i|^2\ran &\sim&  {16\ep^2t\over k^2}A_5(1) \\
\lan |h_{ij}|^2\ran &\sim& 4\ep^2t^3A_5(1) ~,
\eea
where $\Psi$ and $\Phi$ are the Bardeen potentials, $\bm \si$ is the vector
contribution to the shear of the $t=$const. slices and $h_{ij}$ are tensor 
perturbations of the metric.

If it would be solely super horizon perturbations which induce the large
scale CMB anisotropies, this could be translated into a ratio between 
the scalar, vector and tensor contributions to the $C_{\ell}$'s on large 
scales, $\ell\lsim 50$. However, since the main contribution
to the CMB anisotropies is induced at horizon crossing, $x=1$ 
(see below) the above
relations cannot be translated directly and we can just learn that one 
expects, in general, contributions of the same order of magnitude from 
scalar, vector and tensor perturbations.

Finally, we want to discuss in some detail the CMB anisotropies
induced from scalar perturbations. In this case, the 
gravitational perturbation equations (see {\em e.g.} 
\cite{d90,RuthReview}) imply
\be
\Phi +\Psi = -2\ep f_\pi~.
\ee
Especially, if $\Phi$ has a white noise spectrum due to 'compensation'
\cite{MR}, this leads to a $k^{-4}$ spectrum for $\Psi$ and for the 
combination $\Phi-\Psi$ which enters in Eq.~(\ref{dT}). 

This finding is in contradiction with \cite{Joao,Sper}, which predict a
white noise spectrum for $\Psi$, but it is not in conflict with the 
Harrison Zel'dovich spectrum of CMB fluctuations which has been 
obtained numerically in \cite{PST,DZ,Shellard}. This can be seen by the 
following simple argument:
Since topological defects decay inside the horizon, the Bardeen potentials
on sub-horizon scales are dominated by the contribution from dark matter 
and thus roughly constant. The integrated Sachs Wolfe term then contributes
 only up to horizon scales. Therefore, using the fact that for defect models 
$D_g$ and $V$ are much smaller than the Bardeen potentials on 
super-horizon scales (see \cite{MR}), we obtain
\bea
\lefteqn{
(\De T/T)_{\ell}(k)|_{SW} \sim 
	(\Phi -\Psi)(k,x_{dec})j_{\ell}(x_0-x_{dec})}
\nonumber\\ &&
+\int_{x_{dec}}^1({\Phi}'-{\Psi}')(k,x)j_{\ell}(x_0-x)dx~, \label{dT}
\eea
where $x=kt$ and prime stands for derivative w.r.t. $x$. 
The lower boundary of the integrated term roughly cancels the
ordinary Sachs Wolfe  contribution and the upper boundary leads, to
\bea
\lefteqn{\lan|(\De T/T)_{\ell}(k)|^2\ran|_{SW} \sim}\nonumber\\
&&	 {\ep^2\over k^3}
	[3F_5(1)+4F_2(1)+F_3(1)]j_{\ell}^2(x_0) ~,
\eea
a Harrison-Zel'dovich spectrum. The main ingredients for this result are 
the decay of the sources inside the horizon as well as scaling, the rest 
follows for purely dimensional reasons.

\section{Conclusions}
In this paper, we outline a procedure to investigate causal scaling
seed perturbations. We show, that the large number of seed 
correlations, which
determine fully the induced power spectra of dark matter and CMB
photons, can be cast in only five analytic functions with certain well
defined properties. We schematically estimate the large scale CMB
anisotropies induced. However, we are convinced that
the relative amplitudes of large scale CMB anisotropies and the
acoustic peaks as well as the dark matter power spectrum depend on
details of the model and thus scaling causal seeds cannot be ruled out
from studies of specific models. This finding is also confirmed
in\cite{DKLS}. 

Our work is just the beginning of a program to be carried out which
goes in different directions and to which we invite researchers in the
field to participate. Some of the questions which should be explored are the
following: 
\begin{itemize}
\item Are there further general restrictions for the correlation
	functions which have not been mentioned here?
\item Given  the functions $F_1$ to $F_5$ what is the exact expression for
	the $C_\ell$'s and the dark matter power spectrum? What are
	good approximations?
\item  Are there simple conditions which the functions $F_1$ to $F_5$
	have to satisfy in order to lead to power spectra which are in
	agreement with data.
\item Are there  physically plausible causal scaling seed models
	other than topological defects?
\end{itemize}  
\vspace{0.5cm}

{\bf Acknowledgment}\hspace{0.5cm}
This work is partially supported by the Swiss NSF.  Numerical
simulations have been performed at the Centro Svizzero di Calcolo
Scientifico (CSCS).
It is a pleasure to thank P. Ferreira,  M. Sakellariadou and
N. Deruelle for stimulating discussions.

\end{document}